\documentclass[structabstract]{aa} 
\usepackage{graphicx}
\usepackage{color}
\usepackage{txfonts}
\usepackage{longtable}
\usepackage{supertabular}
\usepackage{natbib}
\defcitealias{Dietrich+09}{D09}
\defcitealias{Popesso+07}{P07}
\begin{document}

\definecolor{green}{rgb}{0,0.5,0}
\definecolor{grey}{rgb}{0.4,0.5,0.7}
\newcommand{\msun}{M_{\odot}}
\newcommand{\ks}{\mathrm{km~s}^{-1}}
\newcommand{\rv}{r_{\Delta}}
\newcommand{\mv}{M_{\Delta}}
\newcommand{\rtwo}{r_{200}}
\newcommand{\rtwof}{r_{200}^{\, \sigma}}
\newcommand{\rfive}{r_{500}}
\newcommand{\rtwofive}{r_{2500}}
\newcommand{\rume}{r_{\rm{200,U}}}
\newcommand{\mtwo}{M_{200}}
\newcommand{\ctwo}{c_{200}}
\newcommand{\rhor}{\rho(r)}
\newcommand{\br}{\beta(r)}
\newcommand{\cv}{c_{\Delta}}
\newcommand{\rs}{r_{-2}}
\newcommand{\rh}{r_{\rm H}}
\newcommand{\rb}{r_{\rm B}}
\newcommand{\ri}{r_{\rm I}}
\newcommand{\rc}{r_{\rm c}}
\newcommand{\rn}{r_{\nu}}
\newcommand{\rr}{r_{\rho}}
\newcommand{\ra}{r_{\beta}}
\newcommand{\slos}{\sigma_{\rm{los}}}
\newcommand{\sigv}{\sigma_{\rm{v}}}
\newcommand{\qrr}{Q_{\rm r}(r)}
\newcommand{\nr}{\nu(r)}
\newcommand{\vrf}{v_{{\rm rf}}}

\title{Abell 315: reconciling cluster mass estimates from kinematics,
  X-ray, and lensing \thanks{Based in large part on data collected at
    the ESO VLT (prog. ID 083.A-0930}}

\author{
A. Biviano\inst{\ref{ABi}} 
\and P. Popesso\inst{\ref{PPo}}
\and J. P. Dietrich\inst{\ref{JDi},\ref{PPo}}
\and Y.-Y. Zhang\inst{\ref{YZh}}\thanks{Deceased.}
\and G. Erfanianfar\inst{\ref{PPo},\ref{GEr}}
\and M. Romaniello\inst{\ref{MRo},\ref{PPo}}
\and B. Sartoris\inst{\ref{BSa},\ref{ABi}}
}

\offprints{A. Biviano, biviano@oats.inaf.it}

\institute{
INAF-Osservatorio Astronomico di Trieste, via G. B. Tiepolo 11, 34143, Trieste, Italy\label{ABi} 
\and
Excellence Cluster Universe, Boltzmannstr. 2, 85748 Garching bei M\"unchen, Germany\label{PPo} 
\and
Faculty of Physics, Ludwig-Maximilians-Universit\"at, Scheinerstr. 1, 81679 M\"unchen, Germany\label{JDi} 
\and
Argelander-Institut f\"ur Astronomie, Universit\"at Bonn, Auf dem H\"ugel 71, 53121 Bonn, Germany\label{YZh}
\and
Max Planck Institut f{\"u}r Extraterrestrische Physik Physik, Postfach 1312, 85741 Garching bei M{\"u}nchen, Germany\label{GEr}
\and
European Southern Observatory, Karl-Schwarzschild-Str. 2, 85748 Garching bei M\"unchen, Germany\label{MRo}
\and
Dipartimento di Fisica, Universit\`a degli Studi di Trieste, via G. B. Tiepolo 11, 34143 Trieste, Italy\label{BSa}
}
 
\date{February 10, 2017}

\abstract{Determination of cluster masses is a fundamental tool for
  cosmology. Comparing mass estimates obtained by different probes
  allows to understand possible systematic uncertainties.}{The cluster
  Abell 315 is an interesting test case, since it has been claimed to
  be underluminous in X-ray for its mass (determined via kinematics
  and weak lensing). We have undertaken new spectroscopic observations
  with the aim of improving the cluster mass estimate, using the
  distribution of galaxies in projected phase space.}{We identified
  cluster members in our new spectroscopic sample.  We estimated the
  cluster mass from the projected phase-space distribution of cluster
  members using the \texttt{MAMPOSSt} method.  In doing this estimate
  we took into account the presence of substructures that we were able
  to identify.}{We identify several cluster substructures. The main
  two have an overlapping spatial distribution, suggesting a (past or
  ongoing) collision along the line-of-sight. After accounting for the
  presence of substructures, the mass estimate of Abell 315 from
  kinematics is reduced by a factor 4, down to
  $\mtwo=0.8_{-0.4}^{+0.6} \times 10^{14} \msun$. We also find
  evidence that the cluster mass concentration is unusually low,
  $c_{200} \equiv r_{200}/r_{-2} \lesssim 1$. Using our new estimate
  of $c_{200}$ we revise the weak lensing mass estimate down to
  $\mtwo=1.8_{-0.9}^{+1.7} \times 10^{14} \msun$. Our new mass
  estimates are in agreement with that derived from the cluster X-ray
  luminosity via a scaling relation, $\mtwo=0.9 \pm 0.2 \times 10^{14}
  \msun$.}{Abell 315 no longer belongs to the class of X-ray
  underluminous clusters. Its mass estimate was inflated by the
  presence of an undetected subcluster in collision with the main
  cluster. Whether the presence of undetected line-of-sight
    structures can be a general explanation for all X-ray
  underluminous clusters remains to be explored using a statistically
  significant sample.}


\titlerunning{Reconciling cluster mass estimates}
\authorrunning{A. Biviano et al.}

\maketitle

\section{Introduction}
\label{s:intro}
Accurate and precise determination of galaxy cluster masses is of
crucial importance for cosmological studies
\citep[e.g.,][]{Sartoris+12,Sartoris+16}. Cluster masses can be determined from
scaling relations with other cluster properties \citep[see,
  e.g.,][]{KB12}, such as the X-ray luminosity \citep[$L_X$; see,
  e.g.,][]{Popesso+05,Rykoff+08} and temperature \citep[$T_X$; see,
  e.g.,][]{APP05}, the optical or near-infrared luminosity
\citep[e.g.,][]{Popesso+05,Mulroy+14}, the velocity dispersion and
velocity distribution of member galaxies
\citep[e.g.,][]{Munari+13,Ntampaka+15}, and the Sunyaev-Zel'dovich
signal \citep[e.g.,][]{SEM15}. Direct measurements of cluster masses
can be obtained by assuming hydrostatic equilibrium of the X-ray
emitting intra-cluster gas \citep[e.g.,][]{Rasia+06}, by the
measurement of gravitational lensing shear and magnification
\citep[e.g.,][]{Umetsu+14}, and by the analysis of projected
phase-space distribution of cluster galaxies \citep[see, e.g., the
  review by][and references therein]{Biviano08}, the so-called
'kinematic' mass estimate.

All these methods suffer from possible systematics, arising both from
observational biases, and from violating the assumptions on which the
theoretical derivation of the system mass is based.  X-ray mass
estimates can be biased by gas bulk motions and the complex thermal
structure of the X-ray emitting gas \citep{Rasia+06}, lensing mass
estimates by the
unknown source redshift ($z$) distribution (but not
for low-$z$ clusters) and the assumed concentration of the mass
distribution \citep{Hoekstra+15}. Triaxiality \citep{CK07},
miscentering \citep{Johnston+07}, and substructures can affect both
lensing mass estimates \citep{Giocoli+14}, and kinematic mass
determinations \citep{Biviano+06,MBB13}.

A renewed interest in this topic has come from the puzzling
discrepancy between the values of the cosmological parameters inferred
from cluster counts in the Planck survey and from the primary cosmic
microwave background
anisotropies \citep{Planck13-XX}. A mass bias of 40\% has been
suggested to put the two measurements into
agreement. \citet{vonderLinden+14} found the X-ray based Planck
cluster mass estimates to be biased low by 30\% compared to
weak-lensing mass estimates. Their result might not however apply in
general. Other studies have found good \citep[e.g.,][]{Israel+14,Smith+16}, if
not excellent \citep[e.g.,][]{Umetsu+12} agreement between lensing and
X-ray mass estimates of cluster masses. The comparison of mass
estimates from kinematics, with those from lensing and X-ray, have shown
excellent agreement in some cases \citep[e.g.,][]{Biviano+13}, and serious
discrepancies in others \citep[e.g.,][]{Guennou+14}.

The fact that for some clusters different techniques lead to
consistent mass estimates, and for some they do not, might be related
to the dynamical status of these clusters. \citet[][P07
  hereafter]{Popesso+07} claimed the existence of a class of X-ray
underluminous clusters, which would explain the matching discrepancies
between cluster samples extracted from X-ray and from optical surveys
\citep{Donahue+02,Gilbank+04,Basilakos+04,Sadibekova+15}. The matching
appears to be better between cluster samples extracted from optical
and from Sunyaev-Zel'dovich (SZ) surveys \citep{Rozo+15}. Merging clusters
may account for the poor matching between optical and X-ray detected
clusters.  In fact, in merging clusters the peak of the mass
distribution is offset from the peak of the X-ray emission, as seen in
the Bullet cluster \citep{Markevitch+02}, but not from the peak of the
SZ signal \citep{Zhang+14}. Moreover, X-ray cluster surveys are biased
in favor of high-central density, cool-core clusters
\citep{Eckert+11}, and mergers can disrupt a cluster cool-core and
reduce the concentration of diffuse baryons relative to that of the
dark matter \citep{RBL96,Burns+08,Poole+08}.

\citet{Bower+97} argued that low-$L_X$ clusters of high richness and
velocity dispersion ($\sigv$) are systems of galaxies embedded in
large-scale filaments oriented along the
line-of-sight. \citetalias{Popesso+07} noted that these clusters
(which they called 'AXU' for 'Abell X-ray underluminous') are
characterized by a relative low density of galaxies near their core
and a higher fraction of blue galaxies, relative to normal X-ray
emitting clusters. These characteristics could suggest line-of-sight
contamination. On the other hand, \citetalias{Popesso+07} were unable
to find dynamical evidence for substructure in excess of what was
found in normal clusters.  Signature for significant mass infall rates
in the external regions of the AXU clusters was found, based on the
shape of their galaxy velocity distribution.

To highlight the nature of the low-$L_X$ or high $\sigv$ of AXU
clusters, \citet[][D09 hereafter]{Dietrich+09} determined the weak
lensing masses of two such clusters, Abell 315 and Abell 1456 (A315
and A1456 hereafter), at $\overline{z}=0.174$ and 0.135, respectively.
\citetalias{Dietrich+09} could only set an upper limit to the weak
lensing mass of A1456, which was significantly below the kinematic
mass estimate, but consistent with the mass predicted from the cluster
$L_X$. The velocity distribution of member galaxies in A1456 was found
to be very skewed or even bimodal, suggestive of a complex dynamical
structure that could have biased the kinematic mass estimate high. The
X-ray underluminous nature of A1456 could therefore be rejected.

\citetalias{Dietrich+09}'s weak lensing mass estimate of A315, on the
other hand, was found to be consistent with the one determined from
kinematics, but $\sim 3$ times larger than the mass expected from the
cluster $L_X$ using the scaling relation of \citet{Rykoff+08}. A315
thus remained a good AXU candidate.

To gain insight into the nature of this cluster, we obtained
almost 500 redshifts for galaxies in the cluster field, of
which $\simeq 200$ are estimated to be cluster members.  In this paper
we present these new data, that we use to investigate the internal
structure of A315, and re-determine its kinematic mass estimate. In
Sect.~\ref{s:data} we describe our data-set, in Sect.~\ref{s:members}
we identify the cluster members, in Sect.~\ref{s:subcl} we search for
the presence of substructures, and in Sect.~\ref{s:mass} we determine
the cluster mass from kinematics. We discuss our results in
Sect.~\ref{s:disc} and provide our conclusions in Sect.~\ref{s:conc}.

We use $H_0=70$ km~s$^{-1}$~Mpc$^{-1}$, $\Omega_0=0.3$,
$\Omega_\Lambda=0.7$ throughout this paper.  In this cosmology, at the
cluster mean redshift, $\overline{z}=0.174$, 1 arcmin corresponds to
0.178 Mpc.  All errors are quoted at the 68\% confidence level.

\section{The data-set}
\label{s:data}
Abell 315 was observed at the ESO VLT with VIMOS
\citep{LeFevre+03}. The VIMOS data were acquired using 8 separate
pointings, plus 2 additional pointings required to provide the needed
redundancy within the central region and to cover the gaps between the
VIMOS quadrants. Each mask was observed for 1.5 hours, for a total of 15
hours exposure time. The HR-Blue grism was used, covering the spectral
range 415--620 nm with a resolution $\mathrm{R} \sim 2000$. 
We have reduced the data with the ESO data processing pipeline
v2-9-14\footnote{VLT-MAN-ESO-19500-3355,
  \\ ftp://ftp.eso.org/pub/dfs/pipelines/vimos/vimos-pipeline-manual-7.0.pdf}.
Raw science frames were corrected for bias and flat-field and
calibrated in wavelength according to the standard instrument
calibration
plan\footnote{http://www.eso.org/sci/facilities/paranal/instruments/vimos/doc.html}. Flux
calibration was derived from nightly flux standard star
observations. The flux standard stars themselves were processed
following the same steps as science frames and the resulting response
curve was, then, applied to the processed science spectra. In order to
automatize data processing, we have assembled the pipeline recipes in
a Reflex workflow \citep{Freudling+13}. Redshift estimation has been
performed by cross-correlating the individual observed spectra with
templates of different spectral types from \citet{Polletta+07}.
Templates for ordinary S0, Sa, Sb, Sc, and elliptical galaxies were
used to measure redshifts of relatively low redshift galaxies.  The
cross-correlation is carried out using the \texttt{rvsao} package
\citep[\texttt{xcsao} routine][]{KM98} in the IRAF environment.  The
final sample comprises 479 reliable redshifts in the heliocentric
  rest-frame.

Additional redshifts (in the heliocentric rest-frame) for
galaxies in the cluster area were taken from the SDSS-III
\citep{Eisenstein+11,Ross+14} DR10, 499 in total. There are 32 objects
in common to our spectroscopic sample and the SDSS.  For one of them
there is a substantial difference in the two redshift estimates. 
  The VIMOS redshift estimate is however quite uncertain. It was based
  on a spectrum that looks significantly nosier than the SDSS one,
  possibly because of an imperfect slit centering on the galaxy, due
  to the VIMOS focal plane distortion.  For the remaining 31 we
evaluate a mean redshift difference of $-1.7 \times 10^{-4}$, and a
dispersion of $4.4 \times 10^{-4}$. We use this value and the average
uncertainty of the SDSS redshifts, to estimate an average uncertainty
of $\sim 110 \, \ks$ for the cluster rest-frame velocities of our
VIMOS spectroscopic sample. The VIMOS velocity uncertainty is larger
than the average uncertainty of the SDSS velocities, $\sim 30 \, \ks$,
so we choose the SDSS redshift estimate rather than our own, when both
are available for a given galaxy.

Magnitudes and positions for galaxies in the cluster field were gathered from
the SDSS DR10.

\begin{figure}
\begin{center}
\begin{minipage}{0.5\textwidth}
\resizebox{\hsize}{!}{\includegraphics{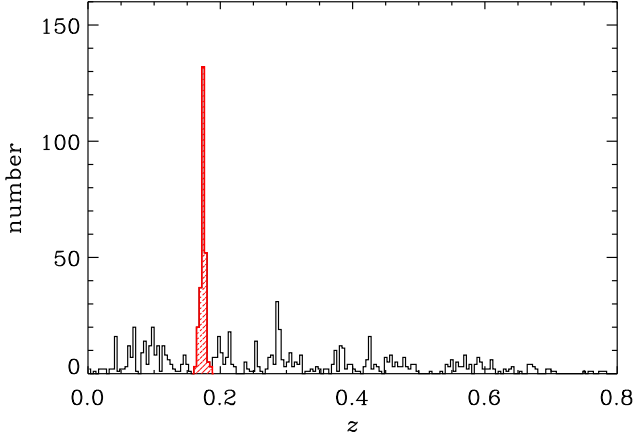}}
\end{minipage}
\end{center}
\caption{Histogram of redshifts in the cluster area. The red, hatched
  histogram shows galaxies with redshifts within $\pm 0.016$ of
  $\overline{z}=0.174$, the mean cluster redshift according to
  \citetalias{Popesso+07}.}
\label{f:zhisto}
\end{figure}

In total our sample contains 946 galaxies with at least one redshift
estimate in the cluster field, over an area of $1\degr 12\arcmin
\times 45\arcmin$. The $z$-distribution of all galaxies in our
spectroscopic sample is shown in Fig.~\ref{f:zhisto}. There is a
prominent peak at the mean cluster redshift, $\overline{z}=0.174$
\citepalias{Popesso+07}.

The spectroscopic sample is presented in Table~\ref{t:zdata}. In
Col.(1) we list a galaxy identification number, in Cols.(2) and (3)
the galaxy right ascension and declination (J2000), in Cols.(4) and
(5) the redshift estimate from SDSS, and from our VIMOS observations
resp., when available, and finally in Col.(6) we flag cluster members
(for the determination of cluster membership see
Sect.~\ref{s:members}), in Col.(7) we flag members in
  substructures identified by the DSb technique (see
  Sect.~\ref{s:subcl} and Appendix~\ref{a:dstest}). In Col.(8) we list
  the probability of a member in the virial region of the cluster, and
  outside DSb-type substructures, to belong to the KMM-main subcluster
  (see Sect.~\ref{s:subcl}).

\begin{table*}
\centering
\caption{The spectroscopic data-set}
\label{t:zdata}
\begin{tabular}{rrrrrccr}
\hline 
\\[-0.2cm]
Id  &  $\alpha_{\mathrm{J2000}}$ & $\delta_{\mathrm{J2000}}$ & $z_{\mathrm{SDSS}}$ & $z_{\mathrm{VIMOS}}$ & Member & Subst & Prob \\[0.15cm]
\hline
\\[-0.2cm]
     2 & $2^{\mathrm{h}} 07^{\mathrm{m}} 36\fs01$ &  $-0\degr 59\arcmin04\farcs7$ &   $0.6056$ &  ---     & ---   & ---  &   --- \\[0.15cm]
   164 & $2^{\mathrm{h}} 07^{\mathrm{m}} 40\fs54$ &  $-1\degr 10\arcmin43\farcs7$ &     ---    & $0.1768$  & M  & ---  &   --- \\[0.15cm]
   328 & $2^{\mathrm{h}} 07^{\mathrm{m}} 44\fs40$ &  $-0\degr 38\arcmin45\farcs3$ &   $0.1748$ & ---      & M  & ---  &   --- \\[0.15cm]
  3272 & $2^{\mathrm{h}} 09^{\mathrm{m}} 06\fs76$ &  $-0\degr 59\arcmin41\farcs5$ &   $0.3732$ & $0.3719$ &  ---  & ---  &  --- \\[0.15cm]
  3664 & $2^{\mathrm{h}} 09^{\mathrm{m}} 53\fs21$ &  $-1\degr 00\arcmin46\farcs0$ &  ---       & $0.1734$ & M  & ---  &  $0.98$ \\[0.15cm]
  3667 & $2^{\mathrm{h}} 10^{\mathrm{m}} 00\fs91$ &  $-0\degr 59\arcmin12\farcs0$ &   $0.1701$ &  ---     & M  & ---  &  $0.07$ \\[0.15cm]
  6437 & $2^{\mathrm{h}} 10^{\mathrm{m}} 35\fs72$ &  $-0\degr 50\arcmin45\farcs6$ &   $0.1785$ &  ---     & M  & S & --- \\[0.15cm]
\hline
\end{tabular}
\tablefoot{The average uncertainties in the VIMOS and SDSS,
    redshifts are $3.7 \times 10^{-4}$ and $1.0 \times 10^{-4}$,
    resp. An 'M' in the 'Member' column identifies cluster members
    (identified as described in Sect.~\ref{s:members}), and an 'S' in the
    'Subst' column identifies galaxies belonging to DSb-type
    substructures (see Sect.~\ref{s:subcl} and
    Appendix~\ref{a:dstest}). The 'Prob' column lists probabilities of
    belonging to the KMM-main subcluster (see Sect.~\ref{s:subcl}).
  Only a portion of the Table is shown here, the full Table is
  available in the electronic version of this paper.}
\end{table*}

\section{Cluster membership}
\label{s:members}
To define which galaxies are members of the cluster we use their
location in projected phase-space $R, \vrf$, where $R$ is the
projected (resp. 3D) radial distance from the cluster center (that we
need to identify) and $\vrf \equiv c \,
(z-\overline{z})/(1+\overline{z})$, is the rest-frame velocity and
$\overline{z}$ is the mean cluster redshift.

Following \citet{Beers+91} and \citet{Girardi+93} we first identify the cluster
main peak in redshift space, by selecting the 252 galaxies with
rest-frame velocities in the range $\pm 4000 \, \ks$, that is within
$\pm 0.016$ of the mean cluster redshift (see Fig.~\ref{f:zhisto}).

\begin{figure}
\begin{center}
\begin{minipage}{0.5\textwidth}
\resizebox{\hsize}{!}{\includegraphics{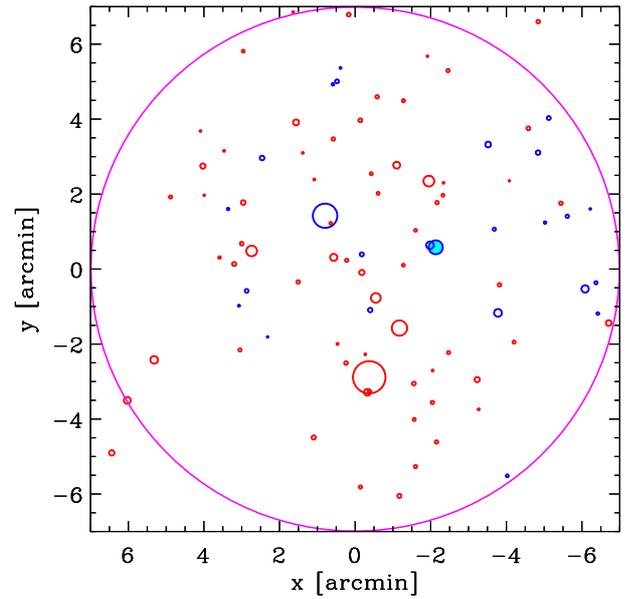}}
\end{minipage}
\end{center}
\caption{The positions of the cluster members with respect to the peak
  of their projected number density (the center is at
  $\alpha_{\mathrm{J2000}}=2^{\mathrm{h}}10^{\mathrm{m}}15\fs0, \;
  \delta_{\mathrm{J2000}}=-1\degr 2\arcmin 31\farcs$0). North is up
  and East is to the left. Galaxy positions are indicated by circles
  with sizes proportional to $1/(m_R-16.5)$, where $m_R$ are the
  galaxy red apparent magnitudes. Red (resp. blue) circles identify
  galaxies with $\vrf \geq -677 \, \ks$ (resp. $<-677 \, \ks$), a
  limit that separates galaxies in the KMM-main subcluster from
galaxies in the KMM-sub subcluster (see Fig.~\ref{f:vkmm} in
Sect.~\ref{s:subcl}). The galaxy selected by \citetalias{Dietrich+09}
as the cluster center is indicated by a blue, cyan-filled, circle at
$\{x,y\}=\{-2.1,0.7\}$. The purple circle has a radius of 1.24 Mpc and
it indicates the cluster virial region (see text).}
\label{f:nobcg}
\end{figure}

\begin{figure}
\begin{center}
\begin{minipage}{0.5\textwidth}
\resizebox{\hsize}{!}{\includegraphics{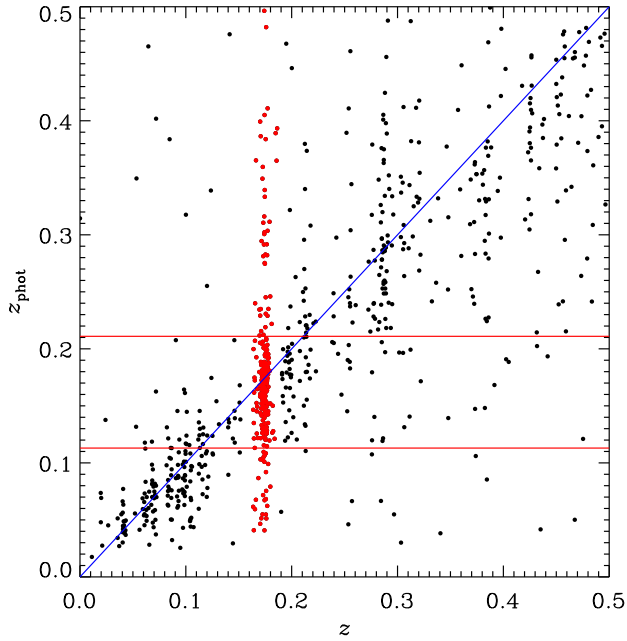}}
\end{minipage}
\end{center}
\caption{Photometric vs. spectroscopic redshift estimates for galaxies
  in the cluster area. Red dots identify galaxies in the main redshift
  peak of Fig.~\ref{f:zhisto}. The blue line represents the
  $z_{\mathrm{phot}}=z$ identity.  The two horizontal red lines
  represent the $z_{\mathrm{phot}}$ limits that we adopt to define cluster
  members for galaxies without $z$.}
\label{f:zpz}
\end{figure}

To define the center of the cluster we cannot rely on the peak of the
X-ray emission, because of poor photon statistics
\citepalias{Dietrich+09}. \citetalias{Dietrich+09} noticed that
the weak lensing peak of A315 was close to a local galaxy overdensity,
and we chose the brightest galaxy of this overdensity as the cluster
center. However, this galaxy does not appear to be the brightest
cluster galaxy, as can be seen in Fig.~\ref{f:nobcg}. In
  this Figure we plot
the cluster members (as defined below) as circles with sizes
proportional to $1/(m_R-16.5)$, where $m_R$ are the galaxy red
apparent magnitudes. Red (resp. blue) circles identify galaxies with
$\vrf \geq -621 \, \ks$ (resp. $<-621 \, \ks$), a limit that separates
galaxies in the KMM-main subcluster from galaxies in the KMM-sub
subcluster (see Fig.~\ref{f:vkmm} in Sect.~\ref{s:subcl}). The galaxy
selected as the cluster center by \citetalias{Dietrich+09} is
part of the KMM-sub subcluster (that we identify in
Sect.~\ref{s:subcl}) and is not the brightest cluster galaxy in the
central cluster region.

Since we can define the cluster center neither from its X-ray emission
nor from the position of a dominant galaxy, we use as a center the
peak of the projected number density of cluster galaxies, that we
determine as follows. We consider the 2D projected spatial density
distribution of cluster members, after correcting our
spectroscopic sample for spatial incompleteness, since some regions
are better covered by spectroscopic observations than others.  To
correct this sample for incompleteness, we rely on a sample with
  homogeneous spatial coverage, that is the sample of photometric
members defined by using the photometric redshifts ($z_{\mathrm{phot}}$)
from SDSS.

In Fig.~\ref{f:zpz} we show the correlation between $z_{\mathrm{phot}}$ and the
spectroscopic redshift $z$, for the 913 galaxies which have both
estimates (we restrict the plot to the redshift range 0--0.5).  We
follow \citet{Knobel+09} and select the $z_{\mathrm{phot}}$ range that
minimizes the metric $\sqrt{(1-P)^2+(1-C)^2}$, where $P, C$ denote
the purity and completeness of the photometric sample of selected
members relative to the sample of 252 spectroscopic members selected
in the main redshift peak. This metric reaches a minimum at $C=0.72,
P=0.59$ for the $z_{\mathrm{phot}}$ range 0.113--0.211, a range we adopt
to select 2327 photometric members.

Of all the selected photometric members, we only consider the 819 brighter than $z_{\mathrm{Petro}} \leq 19.64$
  \citep[corresponding to a luminosity $\simeq 0.13 \, L^{\star}$,
    see][]{MDP09}, a magnitude limit down to which the total
number of galaxies with $z$ is $>1/4$ of the total number of galaxies
with $z_{\mathrm{phot}}$.  We determine the map of spectroscopic
completeness by taking the ratio between the number of spectroscopic
members and the number of photometric members in bins of RA, Dec.
We then assign a completeness value to each galaxy in the
spectroscopic sample and in the chosen magnitude range, according to
the galaxy position.

We have 147 spectroscopic members with $z_{\mathrm{Petro}} \leq 19.64$
and with an assigned spectroscopic completeness $>1/4$, and we use
this sample to construct an adaptive kernel map of the number density
of galaxies in the cluster region, by weighting each galaxy by the
inverse of its completeness value. The resulting map is shown in
Fig.~\ref{f:akmap}, and is centered on the point of maximum
density, located at
$\alpha_{\mathrm{J2000}}=2^{\mathrm{h}}10^{\mathrm{m}}15\fs0, \;
\delta_{\mathrm{J2000}}=-1\degr 2\arcmin 31\farcs$0. This is the
center we adopt for A315. Our adopted center is 0.39 Mpc away
from the position adopted by \citetalias{Dietrich+09}, that was used
as a center for the NFW \citep{NFW97} profile fitting of the weak
lensing map.

Once we have defined the cluster center, we can proceed to a better
identification of the cluster members, by making use not only of the velocity
of galaxies but also of their spatial distribution in the cluster
region. We use the shifting-gapper (SG) algorithm of \citet{Fadda+96}
to identify cluster members in projected phase-space, using a velocity
gap size of 1000 $\ks$, a spatial bin size of 500 kpc, and a minimum of
15 galaxies per spatial bin, as indicated by \citet{Fadda+96}.
We identify 222 cluster members by this method, that is we reject 30
galaxies among those belonging to the main redshift peak. 
The location of the 222 selected members in the cluster area is shown
in Fig.~\ref{f:akmap} and in projected phase-space in
Fig.~\ref{f:rvm}. Hereafter we refer to the sample of 222 cluster
members as the 'Total' sample.

We check our membership definition using the 'Clean' algorithm of
\citet{MBB13}. Using the 'Clean' algorithm the number of
selected members is 208. Differences in the two member selection
algorithms concern only galaxies located at distances $>1$ Mpc from
the center.  In the rest of the paper we present the results based on
the SG membership selection, since the Clean algorithm is based on the
assumption that the cluster mass profile follows a NFW distribution
with a well defined theoretical mass-concentration relation.  The SG
algorithm is instead model-independent. Given that we investigate A315
because of its special properties, we want to avoid biasing the
results by imposing typical characteristics of normal
clusters. Anyway, we checked that the results of this paper are not
significantly dependent on the choice of the membership algorithm.

The mean redshift and velocity dispersion of the cluster members,
evaluated using the biweight \citep{BFG90}, are $\overline{z}=0.1744
\pm 0.0001,$ and $\sigv=603_{-31}^{+29} \ks$ (see also
Table~\ref{t:avsigv}). We use this estimate of $\sigv$ to get a
preliminary estimate of the cluster virial radius\footnote{The radius
  $\rv$ is the radius of a sphere with a mass overdensity $\Delta$
  times the critical density at the cluster redshift. Throughout this
  paper we refer to the $\Delta=200$ radius as the 'virial radius',
  $\rtwo$. Given the cosmological model, the virial mass, $\mtwo$,
  follows directly from $\rtwo$ once the cluster redshift is known, $G
  \, \mtwo \equiv \Delta/2 \, H_z^2 \, \rtwo^3$, where $H_z$ is the
  Hubble constant at the mean cluster redshift.}, $\rtwo$, that we
denote $\rtwof$.  To estimate $\rtwof$ we follow the iterative
procedure of \citet{MBB13}, where we assume an NFW model \citep{NFW97}
for the mass distribution, with a concentration given by the
concentration--mass relation of \citet{MDvdB08}, and we assume the
\citet{ML05b} velocity anisotropy profile with a scale radius
identical to that of the mass profile.  We find $\rtwof=1.24 \pm 0.06$
Mpc. There are 89 members within $\rtwof$.

\section{Substructures}
\label{s:subcl}
\begin{figure}
\begin{center}
\begin{minipage}{0.5\textwidth}
\resizebox{\hsize}{!}{\includegraphics{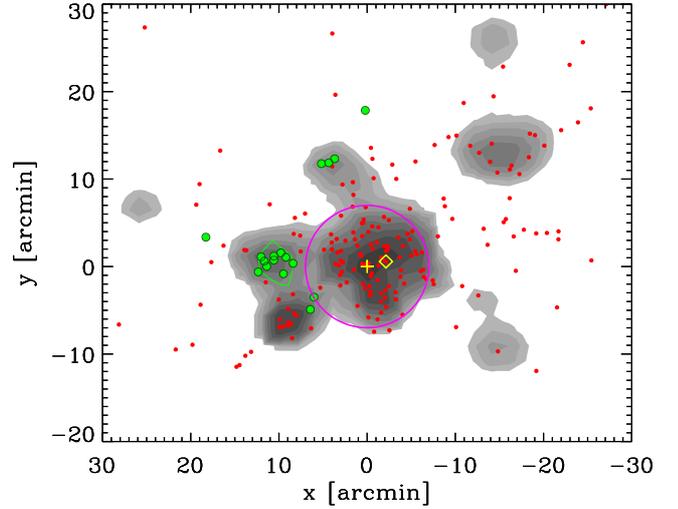}}
\end{minipage}
\end{center}
\caption{Adaptive-kernel map of the number density of cluster members
  with magnitude $z_{\mathrm{Petro}} \leq 19.64$, corrected for
  incompleteness of the spectroscopic sample. Darker shadings indicate
  higher densities, logarithmically spaced. The red dots identify all
  galaxies which are identified as cluster members by the SG algorithm
  (see Sect.~\ref{s:members}). The green dots identify the galaxies
  flagged by the DSb procedure (described in Appendix~\ref{a:dstest})
  as possible members of substructures. The green polygon indicates 10
  of these galaxies that appear to form a compact group (the 'DSb
    group').  The map is centered at the point of maximum projected
  number density of cluster galaxies, as in Fig.~\ref{f:nobcg} (also
  indicated by a yellow plus sign). North is up and East is to the
  left.  The yellow diamond symbol identifies the position of the
  galaxy used as a cluster center in \citetalias{Dietrich+09}. The
  purple circle has a radius of 1.24 Mpc and indicates the cluster
  virial region (see Sect.~\ref{s:members}).  }
\label{f:akmap}
\end{figure}

\begin{figure}
\begin{center}
\begin{minipage}{0.5\textwidth}
\resizebox{\hsize}{!}{\includegraphics{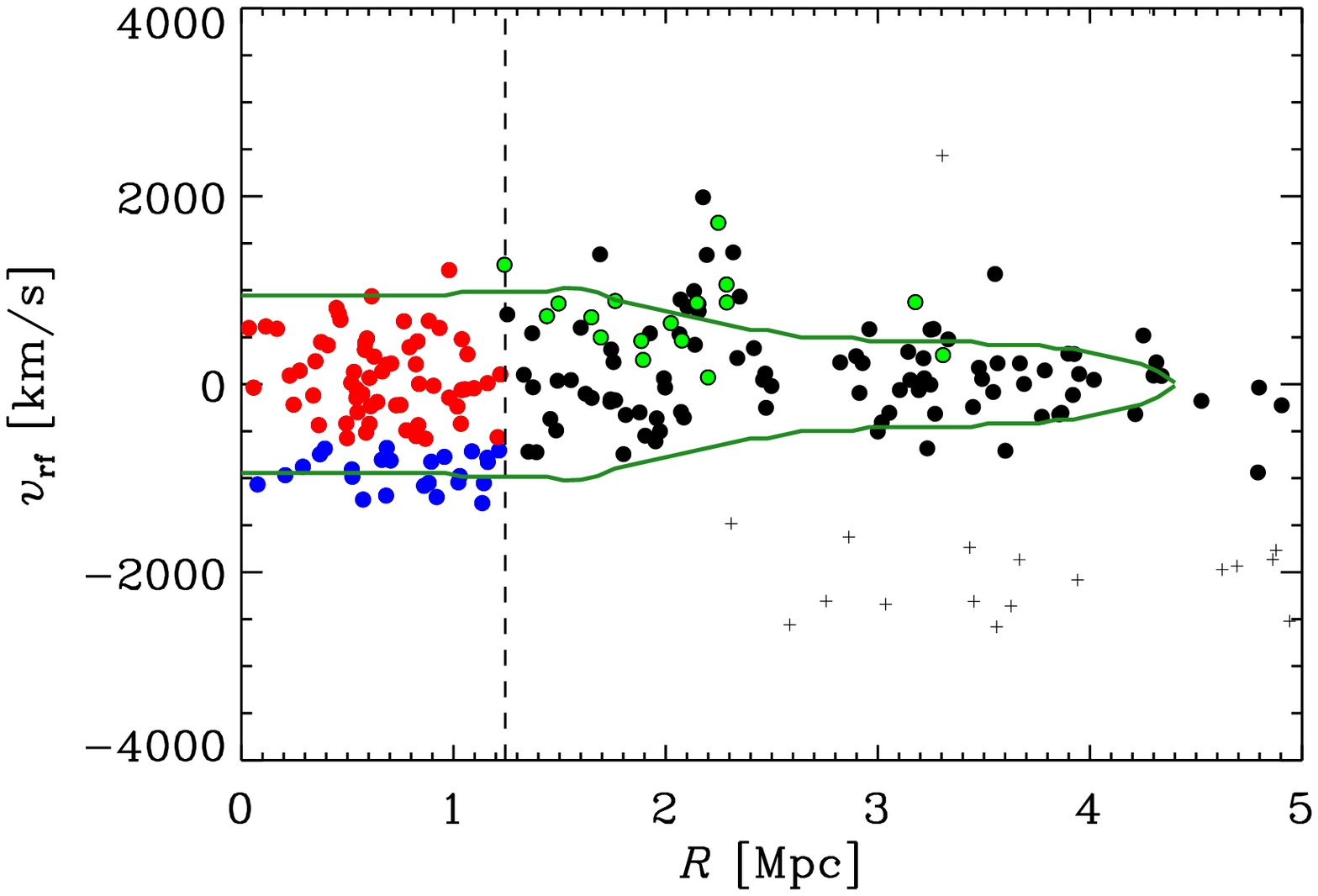}}
\end{minipage}
\end{center}
\caption{The projected phase-space distribution of galaxies in the
  cluster field, $\vrf$ vs. $R$. Crosses and dots represent
  interlopers and cluster members, respectively, identified by the SG
  algorithm of \citet{Fadda+96}. The vertical line is a preliminary
  estimate of the cluster $\rtwo$ ($\rtwof$) based on the global
  estimate of the cluster velocity dispersion (see
    Sect.\ref{s:members}). Members within $\rtwof$ are identified
  with blue (resp. red) dots, if their probability to belong to the
  KMM-sub (resp. KMM-main) subclusters identified by
  the KMM algorithm \citep{McLB88,ABZ94} is $\geq 0.5$ (see text and
  Fig.~\ref{f:vkmm}).  The green colored dots indicate those members
  that are flagged by the DSb procedure (described in
  Appendix~\ref{a:dstest}) as possible members of substructures. The
  green curves represent the Caustics identified by the Caustic
  technique of \citet{DG97} (see Sect~\ref{ss:caustic}).}
\label{f:rvm}
\end{figure}

We consider the presence of substructures in the cluster by using the
test of \citet{DS88}, modified as described in
Appendix~\ref{a:dstest}. This test (DSb test hereafter) looks for
local deviations of the mean velocity and velocity dispersion from the
global cluster values.  We apply the DSb test to the sample of cluster
members defined in Sect.~\ref{s:members}.  In total, 17 members are
flagged for their significant deviation in velocity from the local
mean. Of these, 10 form a compact group in projection (see
Fig.~\ref{f:akmap}), that we call the 'DSb group' hereafter. It has a
mean velocity of $584 \pm 95 \, \ks$ in the cluster rest-frame, and a
velocity dispersion of $282_{-58}^{+72} \, \ks$ (see also
Table~\ref{t:avsigv}), typical of the general population of galaxy
groups \citep[see, e.g., Fig.3 in][]{Ramella+99}. The DSb substructure
galaxies (including the DSb group) are displayed in the
projected phase-space plot of Fig.~\ref{f:rvm}.

After removing the 17 galaxies flagged by the DSb algorithm from the
Total sample, we are left with 205 members, the 'No-DSb' sample
hereafter.

To investigate the presence of additional substructures that remain
undetected by the DSb test, we apply the KMM algorithm
\citep{McLB88,ABZ94} to the distribution of rest-frame velocities of
the remaining 205 cluster members. The KMM algorithm fits a
user-specified number of Gaussian distributions to a data-set, and
returns the probability that the fit by many Gaussians is
significantly better than the fit by a single Gaussian.  Each Gaussian fit
corresponds to a putative substructure of the cluster. The algorithm
also returns the probability for each galaxy to belong to any of these
substructures.  Cluster velocity distributions are known to resemble
Gaussians \citep[e.g.,][]{Girardi+93}, but not when substructures are
present \citep[e.g.,][]{Beers+91}, in which case the decomposition of
the velocity distribution into multiple Gaussians provides a more
appropriate fit to the data \citep[e.g.,][]{Boschin+08}.

\begin{table}
\centering
\caption{Mean velocities and velocity dispersions}
\label{t:avsigv}
\begin{tabular}{lrrrr}
\hline 
\\[-0.2cm]
Sample             &  $N$ & $\overline{v}$~~~~ & $\sigv$~~~ & TI~~  \\[0.15cm]
                   &      & $\ks$ & $\ks$ & \\
\hline
\\[-0.2cm]
Total             & 222  &    $0 \pm 40$ & $603_{-31}^{+29}$ & 1.07 \\[0.15cm]
DSb group  &  10  &  $584 \pm 95$ & $282_{-58}^{+72}$ & --   \\[0.15cm]
No-DSb   & 205  &  $-60 \pm 40$ & $573_{-29}^{+28}$ & 1.05 \\[0.15cm]
Inner     &  88  & $-205 \pm 66$ & $613_{-44}^{+48}$ & 0.88 \\[0.15cm]
KMM-main    &  63  &   $73 \pm 56$ & $441_{-38}^{+41}$ & 0.93 \\[0.15cm]
KMM-sub    &  25  & $-924 \pm 39$ & $189_{-25}^{+29}$ & 0.94 \\[0.15cm]
Outer     & 117  &   $28 \pm 46$ & $503_{-32}^{+34}$ & 1.02 \\[0.15cm]
\hline
\end{tabular}
\tablefoot{Values of the rest-frame mean velocity, the line-of-sight
  velocity dispersion, and the Tail Index (TI; see text) of the
  cluster as a whole and split in several subsamples, and of the
  detected substructures.  The mean velocity and the velocity
  dispersion are computed using the robust biweight estimator
  \citep{BFG90}.  $N$ is the number of objects in each sample.  There
  are 17 DSb galaxies, of which 10 form a group, indicated as 'DSb
  group' in the Table.  The 'No-DSb' sample is obtained from the
  'Total' after removal of the 17 DSb galaxies.  'Inner' and 'Outer'
  are subsamples of 'No-DSb', separated in radial distance by the
  value of $\rtwof$.  `KMM-main' and `KMM-sub' are
  subsamples of 'Inner', identified with the KMM algorithm, and
  separated in velocity space by the value $-621 \, \ks$.  }
\end{table}

We consider the No-DSb sample, and the two subsamples of 88 members
within $R \leq \rtwof$and the 117 members outside ('Inner' and 'Outer'
subsamples, hereafter).  The KMM test indicates that the velocity
distributions of both the No-DSb sample and the outer subsample are
not significantly better fit with 2 Gaussians than with a single one.
On the other hand, a 2-Gaussians fit to the velocity distribution of
the inner subsample is significantly better than a single-Gaussian
fit, with a probability of 0.05.

\begin{figure}
\begin{center}
\begin{minipage}{0.5\textwidth}
\resizebox{\hsize}{!}{\includegraphics{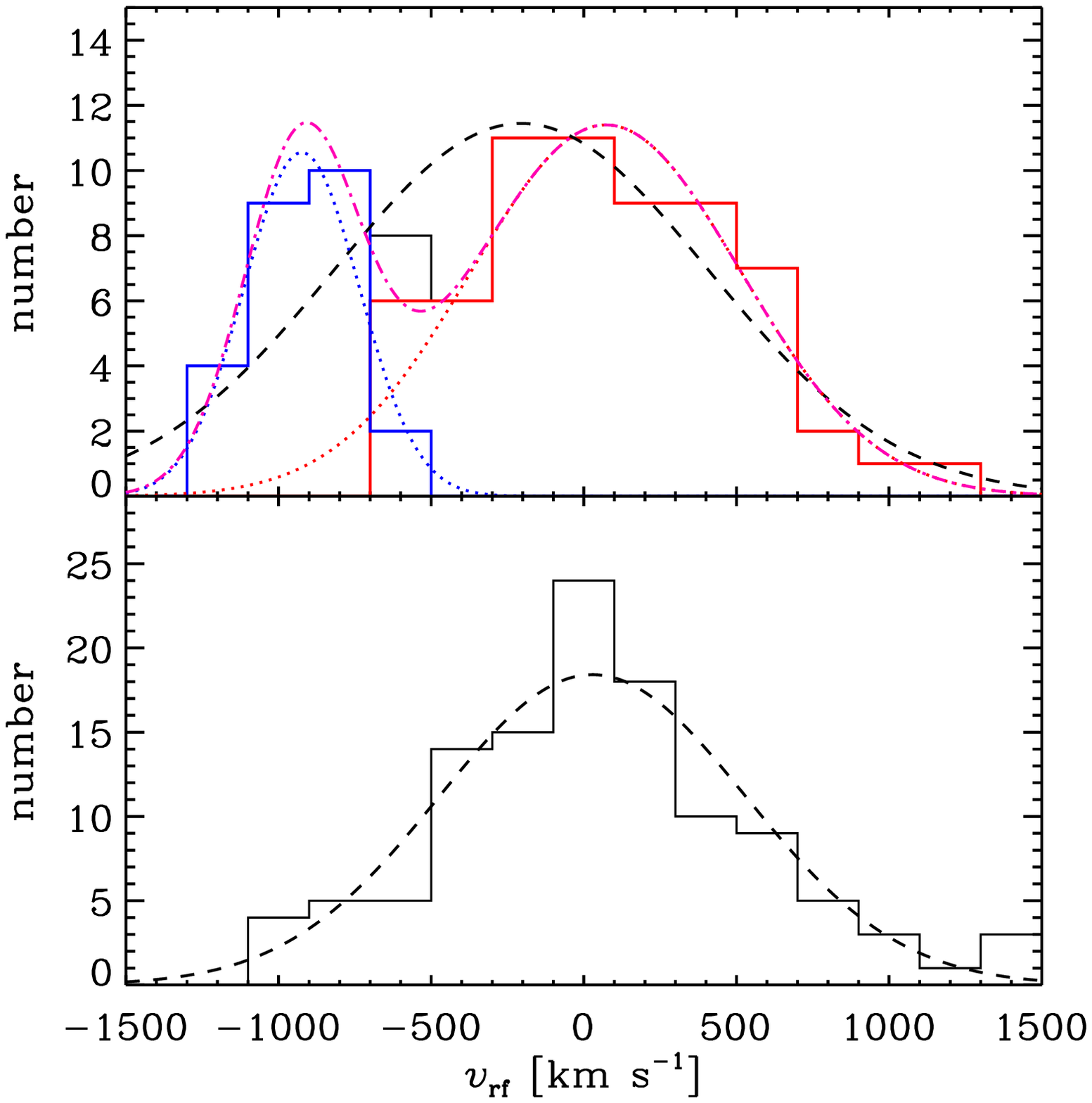}}
\end{minipage}
\end{center}
\caption{The velocity distribution of cluster members (after excluding
  galaxies flagged by the DSb substructure analysis).  Top panel:
  members within $\rtwof=1.24$ Mpc. The blue and red histograms
  identify the KMM partitions (namely, members with probability $\geq
  0.5$ and, respectively, $<0.5$ to belong to the low-velocity group,
  KMM-sub). The dotted (blue and red) curves are the two Gaussians with
  mean and velocity dispersions obtained from the subsamples of the
  same colors. The dash-dotted magenta curve is the sum of the two
  Gaussians. The black dashed curve is the Gaussian with mean and
  velocity dispersion obtained from the full sample. Bottom panel:
  members outside $\rtwof=1.24$ Mpc (histogram). The black dashed curve
  is the Gaussian with mean and velocity dispersion obtained from the
  full sample.}
\label{f:vkmm}
\end{figure}

We show the velocity distribution of the inner subsample, separated
according to the two KMM partitions, in the upper panel of
Fig.~\ref{f:vkmm}, and the velocity distribution of the outer
subsample in the lower panel of the same figure.  We also show the
Gaussians with averages and dispersions obtained from the biweight
estimator \citep[e.g.,][]{BFG90} applied to the different
distributions. In the projected phase-space plot of Fig.~\ref{f:rvm}
we use red and blue dots to distinguish the two groups identified by
the KMM algorithm in the inner sample.

\begin{figure}
\begin{center}
\begin{minipage}{0.5\textwidth}
\resizebox{\hsize}{!}{\includegraphics{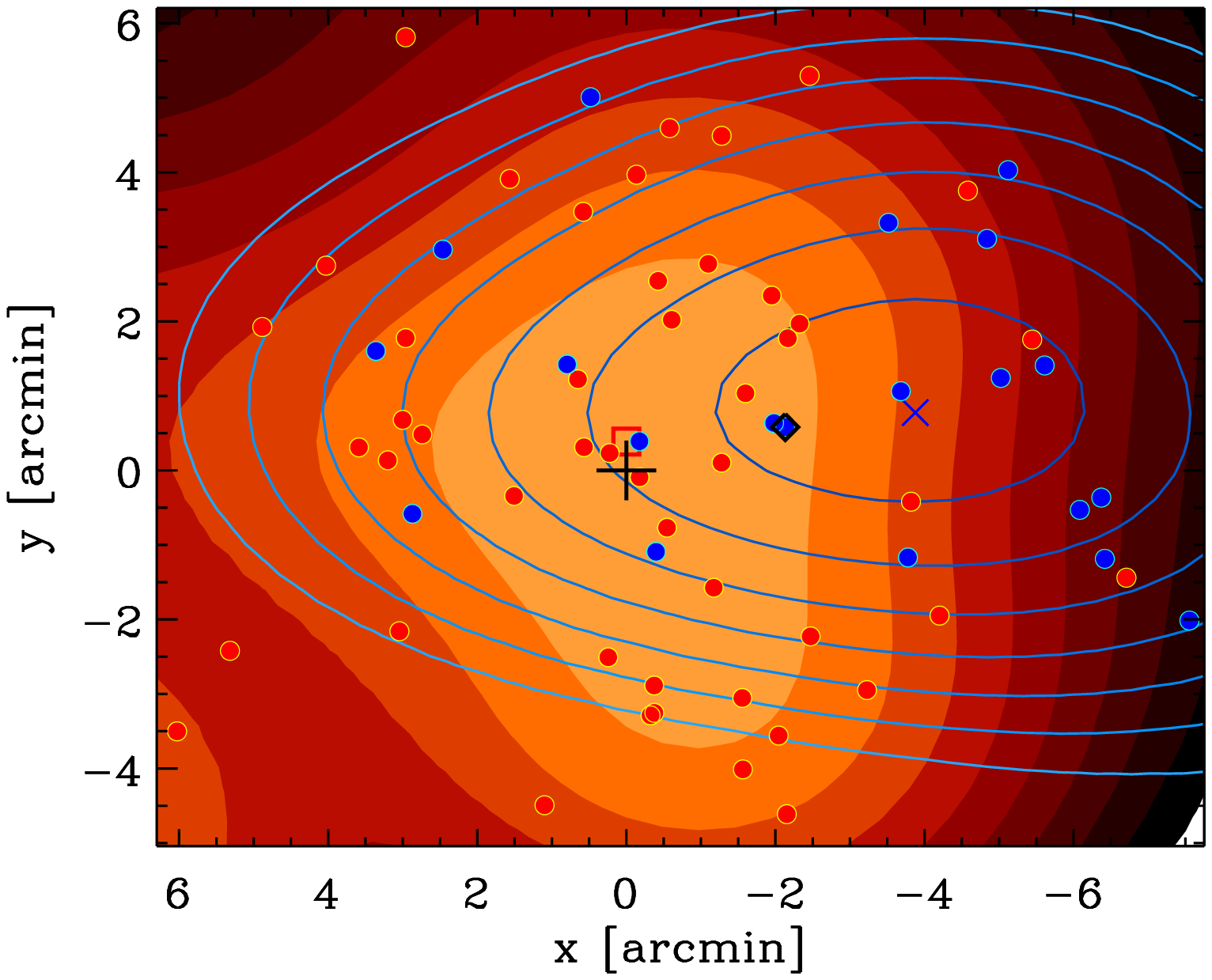}}
\end{minipage}
\end{center}
\caption{Adaptive-kernel maps of the number density of cluster
  members with magnitude $z_{\mathrm{Petro}} \leq 19.64$, corrected for
  incompleteness of the spectroscopic sample. Filled, red-orange
  contours represent the number densities of the KMM-main
  subcluster, open, blue-cyan contours represent the number densities
  of the KMM-sub one. The red square and blue X identify the
  density peaks of the KMM-main and KMM-sub density maps,
  resp.  The contours are logarithmically spaced. The red (blue) dots
  identify member galaxies with velocity $> -621 \, \ks$ (resp. $\leq
  -621 \, \ks$) and are thus more (less) likely to belong to the
    KMM-main subcluster than to the KMM-sub one. The black
  cross identifies our adopted center of A315, from the analysis of
  the adaptive kernel density map of all members. The black diamond
  identifies the center used in \citetalias{Dietrich+09}.}
\label{f:akkmm}
\end{figure}

In Fig.~\ref{f:akkmm} we show the adaptive kernel density map of the
galaxies in the two KMM groups -- restricted to the virial region
where the two groups are defined. As before, we use completeness
weights to construct the map, and we only consider galaxies with
$z_{\mathrm{Petro}} \leq 19.64$ and with an assigned spectroscopic
completeness $>1/4$. We define the KMM-main and KMM-sub subclusters by
considering galaxies of the inner subsample with $\vrf \geq -621 \,
\ks$ and, respectively, and $\vrf <-621 \, \ks$, separated by the
velocity value where the two best-fitting Gaussians intersect in
Fig.~\ref{f:vkmm}.  The density peak of the spatial distribution of
the KMM-main subcluster is nearly coincident (0.07 Mpc
separation) with our adopted center for the whole cluster, as
  expected given that 72\% of the galaxies within $\rtwof$ belong to
the KMM-main subcluster.  The center of the KMM-sub
subcluster, on the other hand, is 0.7 Mpc to the West of the cluster
center. The center used by \citetalias{Dietrich+09} lies at
intermediate distance along the line connecting the two group
centers. The two groups overlap substantially in the projected spatial
distribution, and this overlap is suggestive of a past or
ongoing collision close to the line-of-sight.

In Table~\ref{t:avsigv} we list the values of the average velocities
$\overline{v}$ and velocity dispersions $\sigv$, obtained with the
biweight estimators, for the different samples considered so far. The
removal of the galaxies flagged by the DSb substructure analysis does
not affect the $\overline{v}$ and $\sigv$ values of the whole cluster
significantly. In particular, $\sigv$ decreases by only 5\% when we
remove the 17 DSb-identified galaxies from the total sample. On the
other hand, the $\sigv$ of the Inner sample is significantly larger
than those of the two groups into which it is split by the KMM
algorithm (by 28\% and 69\%).

In the same Table we also list the values of the Tail Index ($TI$) of
the velocity distribution in each sample (except the DSb group, since
10 members are not enough for a reliable estimate of
$TI$). \citet{Beers+91} have suggested the use of $TI$ as a robust
estimator of the shape of the velocity distribution in galaxy
clusters. Values of $TI$ close to unity denote a Gaussian-like
distribution, values $>1$ (resp. $<1$) a distribution with more
  (resp. less) galaxies at large velocity differences than expected
  for a Gaussian (leptokurtic and resp. platikurtic distribution).
\citet{Popesso+07} have found that AXU clusters display on average a
leptokurtic velocity distribution at large radii, with $TI=1.45$, and
interpreted this evidence as suggestive of ongoing infall.

The values we find for the A315 cluster as a whole and for its
different subsamples are not significantly different from unity, not
even for the velocity distribution of members outside the virial
region \citep[see Table 2 in][for the significance levels of the
  $TI$]{BB93}. The velocity distribution within each KMM subcluster is
closer to a Gaussian ($TI=0.93$ and 0.94) than the full velocity
distribution in the virial region ($TI=0.88$). This difference of
  $TI$ values is not significant, but taken at face value it gives
further support to the existence of two subclusters in velocity space.
Had we not excluded the galaxies flagged by the DSb algorithm from our
sample, the $TI$ value of the velocity distribution of the Outer
sample would increase from 1.05 to 1.08, which is also not
significantly different from unity.

\section{The mass estimate}
\label{s:mass}
We proceed to estimate the mass of the cluster by two techniques,
\texttt{MAMPOSSt} \citep{MBB13} and the Caustic \citep{DG97}. In these
estimates, when needed, we take into account the results of the
substructure analysis of Sect.~\ref{s:subcl}. In particular, in \texttt{MAMPOSSt}
we remove the galaxies flagged by the DSb technique, and
we weigh galaxies by their probability of belonging to the KMM-main
subcluster. In the Caustic method we use the KMM-main subcluster
$\sigv$ to select the relevant caustic. 

\subsection{\texttt{MAMPOSSt}}
\label{ss:mamposst}
\begin{figure}
\begin{center}
\begin{minipage}{0.5\textwidth}
\resizebox{\hsize}{!}{\includegraphics{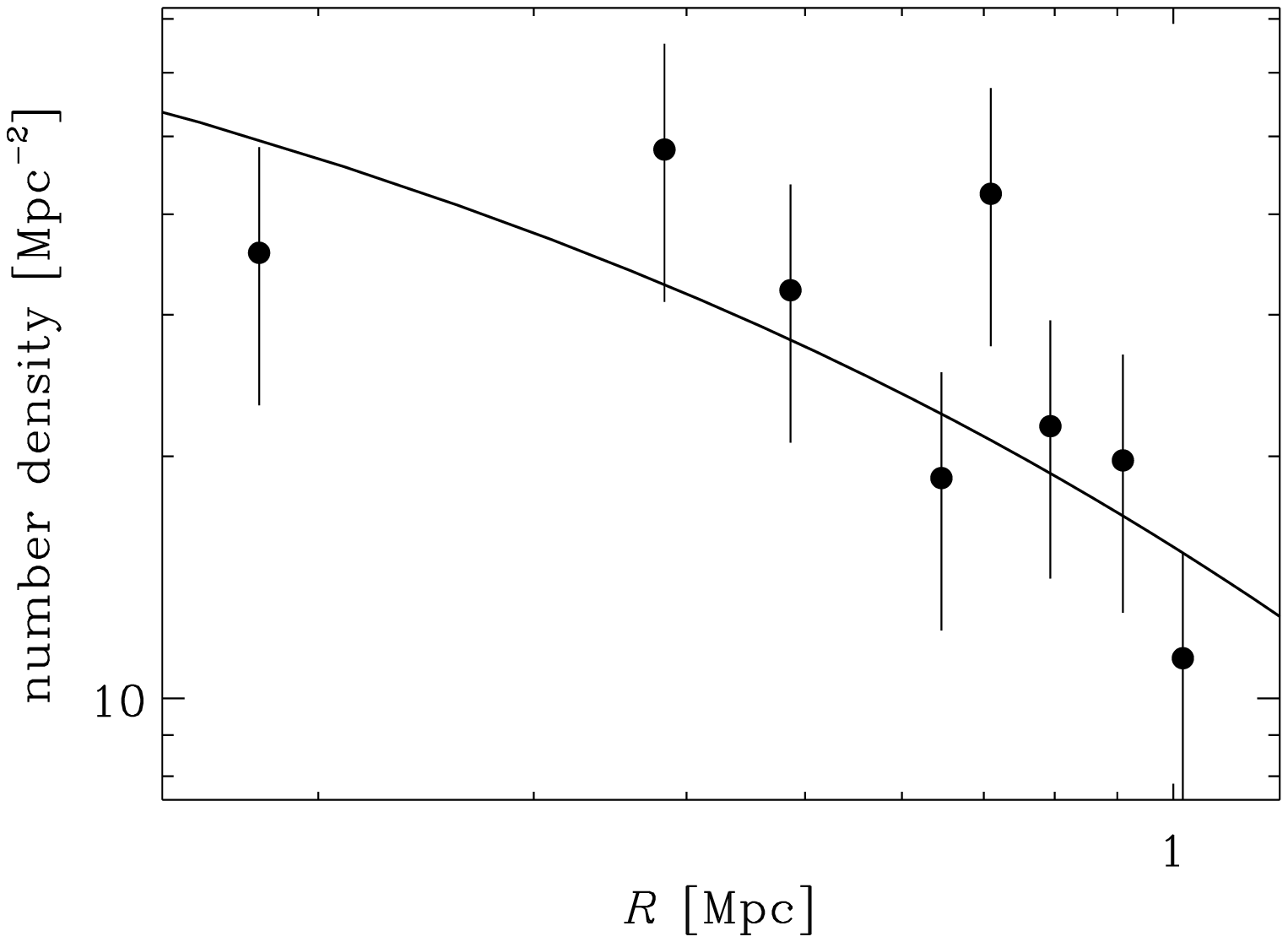}}
\end{minipage}
\end{center}
\caption{The Maximum Likelihood best-fit of a projected NFW model
  \citep{Bartelmann96,NFW97} to the distribution of radial distances
  of the cluster members in the virial region and the binned number
  density profile with 1 $\sigma$ error bars.  Only galaxies with
  $z_{\mathrm{Petro}} \leq 19.64$ and in regions of spectroscopic completeness
  $>1/4$ have been considered, and the sample has been corrected for
  incompleteness.}
\label{f:dprof}
\end{figure}

The \texttt{MAMPOSSt} technique has been developed by \citet{MBB13}.
It determines the best-fit parameters (and their uncertainties) of
models for the mass and velocity anisotropy profile of a system of
collisionless tracers in dynamical equilibrium in a spherical
gravitational potential. To do so, it performs a Maximum Likelihood
analysis of the projected phase-space distribution of the tracers, the
member galaxies of the A315 cluster in our case. It has been tested
with cluster-size halos extracted from cosmological simulations,
by simulating a number of different observational situations.

\begin{figure}
\begin{center}
\begin{minipage}{0.5\textwidth}
\resizebox{\hsize}{!}{\includegraphics{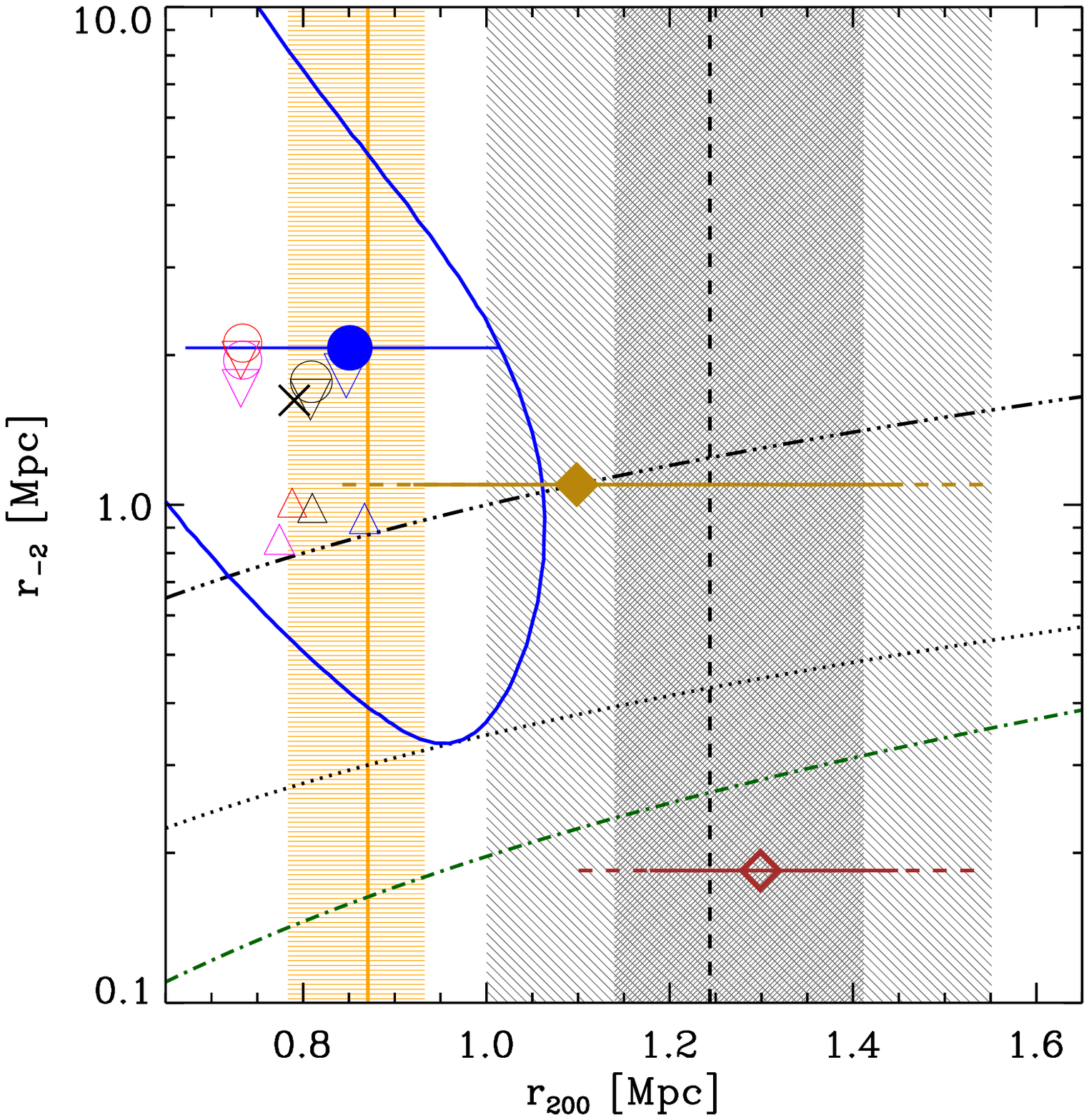}}
\end{minipage}
\end{center}
\caption{Results for the $M(r)$ parameters $\rtwo$ and $\rs$. The blue
  contour indicates the 68\% confidence level on the best-fit values
  obtained in \texttt{MAMPOSSt} (blue dot) for the best-fit models
  NFW+T (see text), after marginalization over the anisotropy
  parameter.  The horizontal solid blue segment indicates the error on
  the best-fit $\rtwo$ value, obtained after marginalization over the
  $\rs$ and the anisotropy parameter.  The best-fit results of other
  models are indicated by the open symbols, triangle, inverted
  triangle, and circle, for the Bur, Her, NFW models, resp., black,
  red, magenta, and blue for the combination with the C, ML, OM, and T
  models, resp. The size of the symbols is proportional to the
  relative likelihood of the models.  The black cross indicates the
  mean $[\rtwo,\rs]$, taking the average over all models. The
  vertical, black dashed line indicate the $\rtwo$ value obtained by
  \citetalias{Dietrich+09} from their kinematic analysis. The
  uncertainties on this value, also taken from
  \citetalias{Dietrich+09}, are indicated by shaded grey regions,
  where the pale grey shading includes both the statistical and the
  systematic uncertainties, while the dark grey shading only includes
  the statistical uncertainty. The vertical orange line and orange
  shading indicate the $\rtwo$ value and uncertainty obtained from the
  cluster $L_X$ \citepalias[from][]{Dietrich+09} using the scaling
  relation of \citet{Rykoff+08}. The open maroon diamond indicates the
  $\rtwo$ value obtained by \citetalias{Dietrich+09} from their
  lensing analysis. The position along the y-axis indicates the $\rs$
  value corresponding to the assumed concentration $\rtwo/\rs$ used in
  \citetalias{Dietrich+09} for the determination of the cluster
  lensing mass.  The statistical and statistical+systematic
  uncertainties on this value are indicated by the maroon solid and
  dashed line, resp.  The filled gold diamond indicates the new
  determination of $\rtwo$ from the lensing analysis applied to the
  same data used in \citetalias{Dietrich+09}, but this time using a
  concentration $\rtwo/\rs=1$.  This value of the concentration is
  used to set the position of the point along the y-axis. The
  dash-dotted green curve is the $\rs$ vs. $\rtwo$ relation derived
  from the concentration-mass relation of \citet{Correa+15-III} at the
  cluster redshift, computed with the code \texttt{COMMAH} \citep[see
    also][]{Correa+15-I,Correa+15-II}. The triple-dot-dashed
  black curve indicates the $\rtwo=\rs$ relation. The dotted
    black curve indicates the $\ctwo=2.9$ relation, namely the highest
    concentration that is still marginally acceptable according to the
    \texttt{MAMPOSSt} dynamical analysis.}
\label{f:rvrs}
\end{figure}

We use \texttt{MAMPOSSt} in the so called ``Split'' mode \citep[see
  Sect. 3.4 in][]{MBB13}, that is we separate the maximum Likelihood
analyses of the spatial and velocity distributions of member galaxies.
We prefer to use \texttt{MAMPOSSt} in the Split mode since our
spectroscopic sample suffers from spatially inhomogeneous
incompleteness,
and while this spatial incompleteness
affects the determination of the number density profile,
it is unlikely to affect the observational determination
of the distribution of velocities. 

To estimate the number density profile we consider the same subsample
of spectroscopically selected members that we used to derive the
adaptive kernel map (Fig.~\ref{f:akmap}), restricted to the virial
region, $R \leq \rtwof$. We fit a projected NFW model
\citep{Bartelmann96,NFW97} to the distribution of radial distances
with a Maximum Likelihood technique, weighting the galaxies by the
inverse of their completeness times their probability of belonging to
the KMM-main subcluster (see Sect.~\ref{s:subcl}). This weighting scheme
is to ensure that we are modeling the KMM-main subcluster density
profile, rather than that of the whole Inner sample of members.
The best-fit model is shown in Fig.~\ref{f:dprof}. The best-fit NFW
scale radius is $r_{\nu}=1.0_{-0.3}^{+0.7}$ Mpc. The uncertainties are
large, but taking the result at face value it suggests a very low
concentration of the galaxy distribution.

We then run \texttt{MAMPOSSt} on the Inner sample of members, by
fixing the $r_{\nu}$ value at its best fit. We prefer to consider only
galaxies within the expected virial region, to avoid including regions
too far from virialization in the analysis. It has in fact been shown
by \citet{MBB13} that $\rtwo$ is the optimal choice for minimizing the
uncertainties in the parameter values obtained by
\texttt{MAMPOSSt}. In calculating the likelihoods of the observed
galaxy velocities, similarly to what we have done in the fit to the
number density profile, we weigh each galaxy in the sample by its
probability of belonging to the KMM-main subcluster. 
  Weighing galaxies by their probabilities of belonging to the
  KMM-main subcluster is a way to account for the contamination by
the KMM-sub subcluster, whose presumed members are assigned
little (or zero) weight. We do not however use completeness as weights
in the \texttt{MAMPOSSt} analysis, since the bias in the observational
selection of spectroscopic targets can easily affect the spatial
distribution, but not the velocity distribution of cluster members.

In \texttt{MAMPOSSt} we search for the best-fit values of three free
parameters,
\begin{enumerate}
\item the virial radius $\rtwo$,
\item the scale radius of the mass distribution, that we choose to
  characterize by $\rs$, the radius at which $\mathrm{d} \log \rho /
  \mathrm{d} \log r =-2$, where $\rho(r)$ is the mass density profile,
\item a parameter that characterizes the velocity anisotropy profile,
  $\br = 1 - \frac{\sigma_\theta^2(r) + \sigma_\phi^2(r)}{
  2\,\sigma_r^2(r)} = 1 - \frac{\sigma_\theta^2(r)}{\sigma_r^2(r)}$,
  where $\sigma_\theta, \sigma_\phi$ are the two tangential
  components, and $\sigma_r$ the radial component, of the velocity
  dispersion, and we assume $\sigma_\theta = \sigma_\phi$.
\end{enumerate}

We consider three models for the mass profile, $M(r)$: 1)
\citet{Burkert95}, 2) \citet{Hernquist90}, and 3) \citet{NFW97} (Bur,
Her, and NFW in the following). They are all characterized by two
parameters, that we convert to $\rtwo$ and $\rs$ when needed
\citep[see][for a detailed description of these models]{Biviano+13}.

We consider four models for the velocity anisotropy profile, $\br$: 1)
a model with constant anisotropy at all radii, that we denote 'C', 2)
the model of \citet{ML05b}, that we denote 'ML', 3) the model of
\citet{Osipkov79} and \citet{Merritt85}, that we denote 'OM', and 4)
the 'T' model used in \citet{Biviano+13}. Using four different
models for $\br$ allows us to evaluate how much our results for $M(r)$
are dependent on the poorly known form of $\br$ in clusters of
galaxies.

\begin{table}
\centering
\caption{\texttt{MAMPOSSt} results}
\label{t:mamp}
\begin{tabular}{lrrrr}
\hline
\\[-0.2cm]
Parameter & NFW+T models & Mean of all models \\[0.15cm]
\hline
\\[-0.2cm]
$\rtwo$ [Mpc] & $0.85_{-0.18}^{+0.16}$ & $0.79 \pm 0.02$ \\[0.15cm]
$\rs$   [Mpc] & $2.1_{-1.0}^{+6.5}$  & $1.6 \pm 0.2$ \\[0.15cm]
$(\sigma_r/\sigma_\theta)_{\infty}$ & $0.7_{-0.3}^{+0.7}$ & $0.8 \pm 0.1$ \\[0.15cm]
\hline
\end{tabular}
\tablefoot{The mean and associated errors have been computed using the
  biweight estimator \citep{BFG90}. The ``mean of all models''
  considers all 12 combinations of 3 models for $M(r)$ and 4 models for
  $\br$, except for the $(\sigma_r/\sigma_\theta^2)_{\infty}$
  parameter, which is only defined for the T $\br$ model.}
\end{table}

The best-fit of \texttt{MAMPOSSt} is obtained for the combination of
the NFW and T models. All other models are statistically acceptable,
at better than the 68\% confidence level.  In Table~\ref{t:mamp} we
give the best-fit values and uncertainties of $\rtwo, \rs$, and the
anisotropy parameter, as well as the mean (and rms) of these same
parameters, obtained by averaging over all the different model
combinations. These values are also plotted in the plane of $\rs$
vs. $\rtwo$ in Fig.~\ref{f:rvrs}. The variance of the values among
different models is substantially smaller than the uncertainties in
the best-fit model, indicating that the results are dominated by the
statistical error, and the precise choice of the $M(r)$ and $\br$
models does not affect our conclusions.

The best-fit $\rtwo$ value found by \texttt{MAMPOSSt},
$\rtwo=0.85_{-0.18}^{+0.16}$, is significantly below our preliminary
estimate, $\rtwof=1.24 \pm 0.06$ Mpc.  This difference is due to the
fact that here we adopt a weighting scheme that effectively forces
\texttt{MAMPOSSt} to consider mostly (if not only) the velocities of
the members of the KMM-main subcluster, while the $\rtwof$ value was
derived from the $\sigv$ estimated using the velocity distribution of
all the cluster members. We repeat our $\sigv$-based estimate of the
virial radius by considering only those galaxies with a probability
$\geq 0.5$ of belonging to the KMM-main subcluster. We find $\rtwof=0.90
\pm 0.09$ Mpc, fully consistent with the \texttt{MAMPOSSt} result. For
comparison, the corresponding value for the KMM-sub subcluster is $0.38
\pm 0.05$ Mpc.

The uncertainty on the \texttt{MAMPOSSt} value of $\rtwo$ is much
larger than that on $\rtwof$. This difference seems strange, given that
\texttt{MAMPOSSt} uses the full velocity distribution, and not only
its $2^{nd}$ moment.  The fact is, the uncertainty in the
$\sigv$-based estimate ($\rtwof$) is obtained by assuming knowledge of
$M(r)$ and $\beta(r)$. The larger uncertainty of the \texttt{MAMPOSSt}
$\rtwo$ estimate is more realistic, as in the \texttt{MAMPOSSt}
procedure we allowed for a much wider range of $M(r)$ and $\beta(r)$
models and parameters. 

The best-fit $\rs$ value obtained by \texttt{MAMPOSSt} is surprisingly
larger than the $\rtwo$ value, implying a concentration $\ctwo \equiv
\rtwo/\rs < 1$, at odds with theoretical expectations
\citep[e.g.,][]{BHHV13,DeBoni+13}. We show in Fig.~\ref{f:rvrs} that
the expected theoretical value of $\rs$ for a cluster this massive at
this redshift is $\lesssim 0.2$ \citep[we use the \texttt{COMMAH}
  routine by][for this
  estimate]{Correa+15-I,Correa+15-II,Correa+15-III}. Hence the
concentration we find is almost an order of magnitude smaller than
expected.

The anisotropy parameter $(\sigma_r/\sigma_\theta)_{\infty}$ has a
best-fit value below unity, characteristic of tangential orbits, but
with large error bars that do not rule out isotropic or even radial
orbits. Tangential orbits are not commonly seen for cluster galaxies
\citep{BP09,WL10,Biviano+13}, but they seem to be more common in
clusters with subclusters \citep{BK04,MBM14}.

\subsection{Caustic}
\label{ss:caustic}
The Caustic method has been developed by \citet{DG97}, and
\citet{Diaferio99} and is a rather simple way to determine the mass
profile of galaxy clusters from the amplitude of the galaxy velocity
distribution at different distances from the cluster center. In
practice, one estimates the density of galaxies in projected
phase-space, and define iso-density contours.  The iso-density contour
that defines 'the Caustic' is chosen by comparing the square amplitude
in velocity space, weighted by the local density of galaxies, to the
$\sigv$ of cluster members in the virial region.  The Caustic method
is supposed to work independently of the presence of substructures,
and does not require the identification of cluster members, if not for
the purpose of estimating the cluster $\sigv$ in the virial
region. Here we determine the Caustic by using all galaxies with
redshifts in the cluster region (not only members, and including
galaxies in substructures), but fixing the cluster $\sigv$ to the
value found for the KMM-main subcluster (see Table~\ref{t:avsigv}). The
Caustic found is shown in Fig.~\ref{f:rvm}.

To convert the Caustic amplitude (along the velocity axis) into a mass
estimate for the cluster, we need to choose a value for the filling
factor ${\cal F}_{\beta}$ \citep[see][for its definition]{Diaferio99}.
Several values have been used so far, ranging from 0.5 to 0.7
\citep{DG97,Serra+11,Geller+13,Gifford+13}. Using
${\cal F}_{\beta}=\{0.5, 0.7\}$ we find
$\rtwo=0.9_{-0.6}^{+0.3}$ Mpc, and $1.0_{-0.6}^{+0.4}$ Mpc, respectively,
where the uncertainties are evaluated following the prescriptions of
\citet{Diaferio99}. Clearly, the statistical error dominates over the
systematic uncertainty in the value of ${\cal F}_{\beta}$. 

The Caustic analysis provides very poor constraints on $\rtwo$
(and therefore the cluster mass), but taken at face value they
are close to those obtained with \texttt{MAMPOSSt} (Sect.~\ref{ss:mamposst})
in particular for ${\cal F}_{\beta}=0.5$.
 
\section{Discussion}
\label{s:disc}
\begin{table}
\centering
\caption{$\mtwo$ estimates}
\label{t:m200}
\begin{tabular}{lcl}
\hline
\\[-0.2cm]
Method & $\mtwo$ & Reference \\
       & [$10^{14} \, \msun$] & \\[0.15cm]
\hline
\\[-0.2cm]
Lensing & $3.0_{-0.8-0.5}^{+1.2+0.7}$ & \citetalias{Dietrich+09} \\[0.15cm]
Virial & $2.7_{-0.7}^{+1.1} \pm 1.0$ & \citetalias{Dietrich+09} \\[0.15cm]
$L_X$ & $0.9_{-0.2}^{+0.2}$ &  \citetalias{Dietrich+09} \\[0.15cm]
\texttt{MAMPOSSt} & $0.8_{-0.7}^{+0.8}$ & This paper \\[0.15cm]
Caustic ${\cal F}_{\beta}=0.5$ & $0.9_{-0.9}^{+1.4}$ & This paper \\[0.15cm]
Caustic ${\cal F}_{\beta}=0.7$ & $1.5_{-1.4}^{+2.4}$ & This paper \\[0.15cm]
Lensing with $\ctwo=1$ & $1.8_{-0.9}^{+1.7}$ & This paper \\[0.15cm]
\hline
\end{tabular}
\tablefoot{Statistical and systematic errors are listed (in this order)
  for the mass estimates of \citetalias{Dietrich+09}.}
\end{table}

In Table~\ref{t:m200} we list the cluster $\mtwo$ values found in this
paper and in \citetalias{Dietrich+09}. Both statistical and systematic
errors are given for the mass estimates of
\citetalias{Dietrich+09}. For the \texttt{MAMPOSSt} mass estimates,
the listed errors include the systematics related to the unknown mass
and velocity anisotropy distributions, since our choice of $M(r)$ and
$\beta(r)$ models has not been restrictive. As for the Caustic mass
estimates, the systematic error is dominated by the choice of ${\cal
  F}_{\beta}$, for which we have considered the two extreme values
generally adopted in the literature.

Our new kinematic estimates of $\mtwo$ are in agreement with the one
obtained from the cluster $L_X$ using the scaling relation of
\citet{Rykoff+08}. On the other hand, our new estimates are
substantially below (by a factor $\sim 3$) the one obtained by the
kinematic analysis of \citetalias{Dietrich+09} which was based on a
sample of 25 cluster members.

Numerical simulations indicate that a bias $>2$ is not unexpected in
kinematic mass estimates based on only $\sim 20$ spectroscopic
members, as it occurs in 25\% of the cases \citep{Biviano+06}.  In
these simulations, the presence of substructures along the
line-of-sight was identified as the main cause of a large bias in the
mass estimate \citep{Biviano+06}. While we could not identify any sign
of subclustering in A315 with a sample of only 25 members, thanks to
our extensive spectroscopic campaign, we have now been able to detect
one small group in the external cluster regions, and, most
importantly, a distinct bimodality in velocity space in the inner
cluster region. This bimodality is due to two subclusters with
an overlapping spatial distribution that suggests they are colliding
or have collided close to the line-of-sight. The $\rtwof$ estimates of
the KMM-main and KMM-sub subclusters imply a mass ratio of
$\sim 0.1$.  Adding the KMM-sub $\mtwo$ to the total cluster
mass estimate therefore does not change our conclusion that our
previous kinematic mass estimate of A315 has been grossly
overestimated.

\begin{figure}
\begin{center}
\begin{minipage}{0.5\textwidth}
\resizebox{\hsize}{!}{\includegraphics{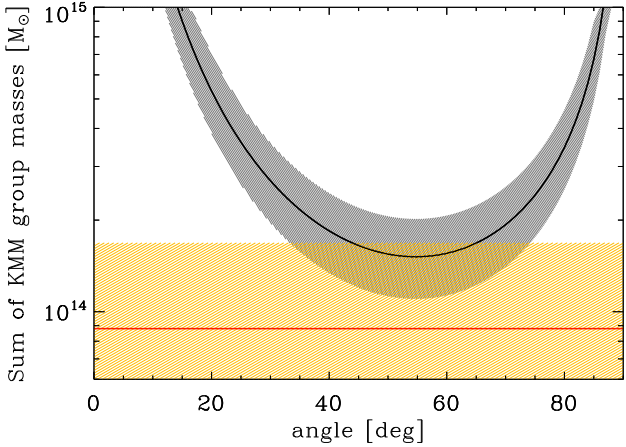}}
\end{minipage}
\end{center}
\caption{The black line shows the requirement for the two KMM
    subclusters to be gravitationally bound, as computed from the
    Newtonian criterion; the grey shading indicates 1 $\sigma$
    uncertainties on this requirement, which depends on the projected
    distance and line-of-sight velocity difference of the two
    subclusters.  The red line indicates the
    sum of the measured masses for these subclusters; the orange
    shading indicates 1 $\sigma$ uncertainties on this sum.  On the
  x-axis, the angle between the axis connecting the KMM-main and
    KMM-sub subclusters is defined with respect to the plane of the
  sky (90 degrees corresponds to a line-of-sight collision).  The
    KMM subclusters are gravitationally bound for an angle between
    $\sim 35$ and $\sim 75$ degrees, if we take into account both the
    uncertainties on our mass estimates and the uncertainties on the
    Newton criterion requirements.}
\label{f:2body}
\end{figure}

If the two subclusters are physically unrelated, and their velocity
difference attributed to different Hubble flows, the smallest
subcluster would lie $\sim 18$ Mpc in the foreground. However,
  the subclusters are unlikely to be completely unrelated, as
we can see by applying the Newtonian criterion for gravitational
binding of the two subclusters \citep[eq.(5) in][]{Beers+91}. To
  apply the Newtonian criterionwe use the difference in the
  subcluster mean line-of-sight velocities ($851 \pm 68 \, \ks$), and
the projected separation between their centers ($0.66 \pm 0.10$
Mpc). We use the \texttt{MAMPOSSt} $\mtwo$ estimate for the
  KMM-main subcluster (see Table~\ref{t:m200}) and 1/10 of this same
estimate for the KMM-sub subcluster. The result is shown in
Fig.~\ref{f:2body} and indicates that a bound solution is acceptable
within the observational uncertainties, for a wide range of values of
the angle between the collision axis and the plane of the sky. The
bound solution is even more likely than our estimate indicates,
because we have used $\mtwo$ masses, and these do not account for the
additional mass within the infall regions of the two subclusters
\citep[the total mass of the system would increase by a factor $\sim
  2$;][]{RGDK13}.

The bound solution does not inform us on whether the two subclusters
are observed before or after their collision. A past collision between
the two subclusters might be invoked to explain the very low
concentration ($\ctwo<1$) observed both for the galaxy and the mass
distribution of the main component of A315. Such a low concentration
is indeed uncommon \citep{LMS04,Budzynski+12} and theoretically
unexpected \citep[e.g.,][]{Correa+15-III}. Observationally, it has
been shown that the radial distribution of galaxies in clusters with
substructures is less concentrated than that of galaxies in relaxed
clusters \citep{Biviano+02}. On the theoretical side, numerical
simulations have shown that the scale radius of the mass distribution
increases after a merger \citep{Hoffman+07}. A low-concentration of
the mass distribution characterizes not fully virialized clusters
\citep{Jing00,Neto+07}.

Could then the low concentration we observe originate from the
  collision with the subcluster identified by the KMM analysis? To
  answer this question, we estimate the probability of a halo of mass
  similar to the mass of A315, to have a concentration $\ctwo<2.9$.
  This is the highest value that is still marginally acceptable
  according to our \texttt{MAMPOSSt} dynamical analysis of A315
  (dotted curve in Fig,~\ref{f:rvrs}). We use the concentration
  distributions of the halos in the Millennium Simulation derived by
  \citet{Neto+07}. More precisely, we consider the lognormal best-fit
  models listed in their Table~1, for the halos in the mass range
  closest to our A315 mass estimate.  While only 1\% of relaxed halos
  have $\ctwo<2.9$, 28\% of unrelaxed halos have such a low
  concentration or lower. This fraction drops to 0.005\% at
  $\ctwo<1$. It then appears that the best-fit concentration value we
  observe is rarely observed in cosmological simulated halos, but not
  when we account for the observational uncertainties and for the
  unrelaxed nature of A315.

The low-concentration of the mass distribution of A315 might also
account for part of the mass overestimate from lensing
\citepalias{Dietrich+09}.  \citetalias{Dietrich+09} treated the NFW
profile as a 1-parameter profile where the concentration follows the
theoretical mass-concentration relation of \citet{Dolag+04}
exactly. At the best fit $\mtwo$ in \citetalias{Dietrich+09} the
concentration used was 7.0. Performing a two-parameter fit, with
  a free concentration parameter is unfortunately not allowed by the
  quality of the \citetalias{Dietrich+09} data. In particular, the low
  total number of galaxies inside the NFW scale radius limits the
  constraining power of this data set. Furthermore, contamination of
  the catalog of lensed galaxies by cluster galaxies dilutes the shear
  signal in a radially-dependent way that is extremely challenging to
  model even for much better quality data than those of
  \citetalias{Dietrich+09}. We therefore repeat the weak lensing
  analysis of \citetalias{Dietrich+09} on the same data and with the
  same technique, but this time forcing $\ctwo=1$ instead. We obtain
  $\mtwo=1.8_{-0.9}^{+1.7} \times 10^{14} \msun$, that brings the
  lensing mass estimate in agreement with the kinematic and X-ray
  estimates within 1 $\sigma$ (see Fig.~\ref{f:rvrs} and
  Table~\ref{t:m200}).

In addition, the lensing mass estimate might be further reduced by
considering that it is derived assuming a spherical NFW profile, while
the cluster mass distribution is elongated along the line-of-sight due
to the two overlapping subclusters
\citep[e.g.,][]{CK07,Dietrich+14}. If the elongation is only due to
the superposition of the two subclusters, we expect the effective axis
ratio of the total mass distribution not to be too far from unity.
However, in low concentration clusters, the mass ratio between
  the best-fitting lensing mass obtained assuming a spherical NFW halo
  and the true mass of an elliptical NFW halo, can be
  $\sim 1.1$ also for a relatively small axis ratio \citep[see
    Fig.~2 in][]{Dietrich+14}.

Due to its dependence on the square of the electron density, X-ray
luminosity-based mass estimates are to a good approximation not
affected by triaxiality.
However, the low mass concentration suggests that A315 might be a
non-cool-core cluster. The mass estimate that one obtains from $L_X$
via a scaling relation obtained for an unbiased cluster sample, is
systematically lower for non-cool-core clusters, by $\sim 25$\%
\citep[see Fig.~3 in][]{Zhang+11}. Indeed, scaling relations with
  core-excised $L_X$ have less dispersion and lower systematics
  than those obtained from the total $L_X$ \citep{MHRJ11}.

The presence of substructure in the velocity distribution of A315 and
its low mass concentration, thus seems to be able to reconcile the
X-ray, lensing, and kinematic cluster mass estimates.  Possibly
the presence of substructures and the low mass concentration
are both the manifestation of the same phenomenon, namely a
collision along the line-of-sight of a poor cluster and a galaxy
group.

In conclusion, our new analysis rules out the X-ray underluminous
nature of A315, just as it was done for A1456 by
\citetalias{Dietrich+09}. These clusters appear X-ray underluminous
because their velocity dispersions are inflated by infalling,
unrelaxed halos -- an interpretation originally given by
\citet{Bower+97} to explain the existence of low-$L_X$ clusters with
high $\sigv$.

A315 and A1456 are however only 2 of 51 AXU clusters in the sample of
\citetalias{Popesso+07}. Both were found to be characterized by a
bimodal velocity distribution when analyzed in detail and with more
spectroscopic data (in the case of A315). Such a velocity distribution
is characterized by low $TI$ values (like the one we obtain for the
Inner sample of A315, see Table~\ref{t:avsigv}), as expected from
  the presence of two kinematically distinct components with a mean
  velocity offset \citep[see, e.g., the case of A85 in the study
    of][]{Beers+91}. However, low $TI$ values are not typical of AXU
  clusters, that {\citetalias{Popesso+07} found instead to have
    velocity distributions characterized by high $TI$ values outside
    the virial region, a feature that remains to be explained. One
possibility is that high $TI$ values are caused by the presence
of high-velocity interlopers that are not removed by the membership
selection procedure, which could fail
for poor statistical samples. More detailed investigations of other
AXU clusters are needed before we can dismiss the existence of
intrinsically X-ray underluminous clusters altogether.

\section{Conclusions}
\label{s:conc}
We re-determine the kinematic mass estimate of the $z=0.174$ cluster
A315, which had previously been identified as being X-ray
underluminous for its kinematic and lensing mass
\citepalias{Popesso+07,Dietrich+09}. Our new kinematic estimate is
based on redshifts for $\sim 200$ cluster members, in part
obtained through our new spectroscopic observations with VIMOS
at the VLT. These are the results of our analysis:
\begin{itemize}
\item We identify previously undetected substructures. In particular,
  the velocity distribution of cluster members in the virial region
  displays a significant bimodality, caused by the projection of two
  distinct subclusters along the line-of-sight.
\item Accounting for these substructures in our kinematic analysis
  \citep[conducted via \texttt{MAMPOSSt} and the Caustic
    method,][resp.]{MBB13,DG97}, leads to a substantial and
  significant reduction of the kinematic mass estimate of
  \citetalias{Dietrich+09}, which was based on 25 members only.  Our
  kinematic mass estimate, $0.8_{-0.4}^{+0.6} \times 10^{14}
  \msun$, is in agreement with the estimate that we obtain from the
  cluster $L_X$ through the scaling relation of \citet{Rykoff+08},
  $0.9 \pm 0.2 \times 10^{14} \msun$.
\item In our dynamical analysis we also determine the cluster mass
  concentration. We find $\ctwo<1$, an unusually low value. We argue
  that this is the effect of a $\sim 1$:10 mass-ratio collision between the
  two subclusters identified in the virial region.
\item Using our estimate of $\ctwo$, we redetermine the weak lensing
  mass of A315 using the same method of
  \citetalias{Dietrich+09}, and we find $\mtwo=1.8_{-0.9}^{+1.7} \,
  \msun$. This mass estimate is 40\% lower than the estimate 
  of \citetalias{Dietrich+09}, which was obtained using a much higher
  concentration, inferred from a theoretical concentration-mass
  relation. Accounting for elongation of the cluster along the
  line-of-sight could further reduce our new lensing mass estimate (by
  $\gtrsim 10$\%).
\item The low-mass concentration we find might suggest that A315 is
  not a cool-core cluster. Its $L_X$ might therefore correspond to a
  slightly higher mass (by $\sim 25$\%) than the one predicted by the
  \citet{Rykoff+08} scaling relation.
\end{itemize}

Our new results dismiss the AXU nature of A315, just as it was done
for A1456 by \citetalias{Dietrich+09}. The A315 $L_X$ no longer
appears too low for its mass. Its lensing mass had been over-estimated
because it was derived assuming a normal mass concentration, rather
than the true, very small one.  The cluster kinematic mass had
previously been over-estimated because of an undetected bimodality in
its velocity distributions. This was also the case of A1456. Both
clusters belong to the category of systems whose velocity dispersions
are inflated by infalling subclusters or groups projected along the
line-of-sight \citep{Bower+97}. Whether line-of-sight projections
  are the only explanation for the nature of AXU clusters is
impossible to say before more candidates are examined with the same
level of detail used for A315. These studies will help quantifying the
biases in cluster mass estimates, a fundamental issue for the use of
clusters as cosmological probes.

\begin{acknowledgements}
We dedicate this work to the memory of our friend and colleague
  Yu-Yin Zhang, whose collaboration we have enjoyed and appreciated
  for several years. We thank the anonynous referee for her/his useful
  comments.  A.B. acknowledges the hospitality of the Excellence
Cluster Universe, and financial support provided by the PRIN INAF
  2014: “Glittering kaleidoscopes in the sky: the multifaceted nature
  and role of Galaxy Clusters”, P.I.: Mario Nonino.
Y.Y.Z. acknowledges support by the German BMWi through the
Verbundforschung under grant 50 OR 1304.  This paper has made use of
data from SDSS-III.  Funding for SDSS-III has been provided by the
Alfred P. Sloan Foundation, the Participating Institutions, the
National Science Foundation, and the U.S. Department of Energy Office
of Science. The SDSS-III web site is http://www.sdss3.org/.  SDSS-III
is managed by the Astrophysical Research Consortium for the
Participating Institutions of the SDSS-III Collaboration including the
University of Arizona, the Brazilian Participation Group, Brookhaven
National Laboratory, Carnegie Mellon University, University of
Florida, the French Participation Group, the German Participation
Group, Harvard University, the Instituto de Astrofisica de Canarias,
the Michigan State/Notre Dame/JINA Participation Group, Johns Hopkins
University, Lawrence Berkeley National Laboratory, Max Planck
Institute for Astrophysics, Max Planck Institute for Extraterrestrial
Physics, New Mexico State University, New York University, Ohio State
University, Pennsylvania State University, University of Portsmouth,
Princeton University, the Spanish Participation Group, University of
Tokyo, University of Utah, Vanderbilt University, University of
Virginia, University of Washington, and Yale University.
\end{acknowledgements}

\bibliography{a315}

\appendix
\section{The modified Dressler \& Shectman (DSb) test}
\label{a:dstest}
The original version of the test looked for these deviations
in all possible groups of 11 neighboring galaxies identified within a
cluster \citep{DS88}.  \citet[][see Appendix A.3 in that
  paper]{Biviano+96} adapted this method to make adaptive-kernel maps
of the quantity $\delta$ that describes the average velocity and
velocity dispersion deviation from the global cluster value.
\citet{Biviano+02} then modified $\delta$ into its two components
$\delta_{v}$ and $\delta_{\sigma}$, that separately measure
the local deviations of the average velocity and velocity dispersion,
respectively. They also introduced the use of the velocity dispersion
profile, in lieu of the total cluster velocity dispersion, as a reference
value for  $\delta_{\sigma}$. 

\begin{figure}
\begin{center}
\begin{minipage}{0.5\textwidth}
\resizebox{\hsize}{!}{\includegraphics{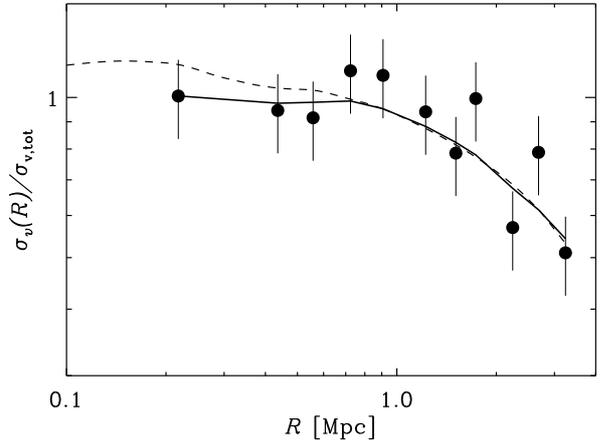}}
\end{minipage}
\end{center}
\caption{The cluster projected velocity dispersion profile, normalized
  by the total velocity dispersion (dots with 1 $\sigma$ error
  bars). The solid curve represent the smoothed profile, adopted in
  the Dressler \& Shectman test.  The dashed curve shows, for
  comparison, the smoothed profile obtained for a sample of nearby
  clusters by \citet{Biviano+02}.}
\label{f:vdp}
\end{figure}

We combine the modifications proposed
by \citet{Biviano+96} and \citet{Biviano+02}. Specifically, we evaluate
the local values of mean velocity and velocity dispersion by
constructing weighted adaptive-kernel density maps of cluster members,
with the weights given by $v$ and $v^2$, and dividing
these maps by the unweighted adaptive-kernel number density map of
cluster members,
\begin{equation}
\delta_{v}(\vec{x})=\sum_{j=1}^N K_j^{(2D)}(\vec{x}) v_j / \sum_{j=1}^N K_j^{(2D)}(\vec{x})
\end{equation}
\begin{equation}
\delta_{\sigma}(\vec{x})=[\sum_{j=1}^N K_j^{(2D)}(\vec{x}) v^2_j / \sum_{j=1}^N K_j^{(2D)}(\vec{x})]^{1/2}-\sigma_{v}(R)
\end{equation}
where $K_j^{(2D)}(\vec{x})$ is the 2D kernel at the position
$\vec{x}$, and $\sigma_{v}(R)$ is the total cluster velocity
dispersion profile, that is the velocity dispersion at a given projected
position $R$, shown in Fig.~\ref{f:vdp}.

The significance of the $\delta_{v}$ and $\delta_{\sigma}$ at any position
$\vec{x}$ are evaluated separately, by bootstrap resamplings \citep{ET86}.

\end{document}